Spectral Risk Measures with an Application to Futures Clearinghouse Variation Margin Requirements

By

John Cotter and Kevin Dowd*

Abstract

This paper applies an AR(1)-GARCH (1, 1) process to detail the conditional distributions of the return distributions for the S&P500, FT100, DAX, Hang Seng, and Nikkei225 futures contracts. It then uses the conditional distribution for these contracts to estimate spectral risk measures, which are coherent risk measures that reflect a user's risk-aversion function. It compares these to more familiar VaR and Expected Shortfall (ES) measures of risk, and also compares the precision and discusses the relative usefulness of each of these risk measures in setting variation margins that incorporate time-varying market conditions. The goodness of fit of the model is confirmed by a variety of backtests.

October 31 2006

* John Cotter is at the Centre for Financial Markets, School of Business, University College Dublin, Carysfort Avenue, Blackrock, Co. Dublin, Ireland; email: john.cotter@ucd.ie. Kevin Dowd is at the Centre for Risk and Insurance Studies, Nottingham University Business School, Jubilee Campus, Nottingham NG8 1BB, UK; email: Kevin.Dowd@nottingham.ac.uk. Cotter's contribution to the study has been supported by a University College Dublin School of Business. We thank seminar participants at CASS Business School, UCD Centre for Financial Markets, Tor Vergata Annual Conference 2005, the joint conference of the European Central Bank and the Federal Reserve Bank of Chicago 2005 on 'Issues Related to Central Counterparty Clearing' and the Deutsche Bundesbank Conference 2005 in celebration of the 80th birthday of Professor Benoît B. Mandelbrot.

# 1 INTRODUCTION

Futures clearinghouses set margins requiring traders to pay a deposit to minimise default risk and acts as counterparty to all trades that take place within its exchanges. This ensures that individual traders do not have to concern themselves with credit risk exposures to other traders, because the clearinghouse assumes all such risk itself. However, it also means that the clearinghouse has to manage this risk, and one way it does so by imposing margin requirements. These consist of an initial margin (deposit) and a variation or daily margin. The initial margin represents the deposit a futures trader must give to a Clearinghouse to initiate a trade whereas in contrast the variation margin is the extra deposit required of the trader once a margin call is made.[1]

In modelling initial margins the focus is typically on extreme confidence levels for extraordinary market events so as to minimise the probability that the associated quantile is exceeded (Longin (1999) and Cotter (2001)). This is equivalent to stress testing for very low probability events. It can also be thought of as requiring unconditional risk modelling over a long forecast horizon, and most previous literature has focused on initial margins modelled unconditionally.[2] The variation margin can thought of as supporting the initial margin after it has been breached to help avoid trader default.[3] But whereas the initial margin is intended to reflect long-run conditions, the variation margin is intended to reflect current market conditions.

---

[1] See Hull (2003) for details of margin requirements for futures. Essentially there are four elements to a margin account: As well as the initial margin; there is the maintenance margin that represents the minimum balance of the margin account that must not be breached; the margin call where an investor is informed that they have to top up their margin account and the variation margin representing the amount that the investor must add to their margin account as a result of the margin call to bring it up to the value of the initial margin. If the trader defaults on paying the variation margin the broker closes out the position by selling the contract.

[2] Many approaches have been followed in setting initial margins. For instance Figlewski (1984), Edwards and Neftci (1988), Warshawsky (1989), Booth et al (1997), and Longin (1999) use different unconditional statistical distributions (Gaussian, historical or extreme value distribution). In contrast, Brennan (1986) proposes an economic model for broker cost minimization where margins are endogenously determined, and Craine (1992) and Day and Lewis (2004) model the distributions of the payoffs to futures traders and the potential losses to the futures clearinghouse in terms of the payoffs to barrier options.

[3] For illustrative purposes we assume that the initial margin is zero and that the variation margin represents the full margin requirement deposit of traders. In reality the variation margin represents the extra funds required to bring the traders deposit back to the value of the initial margin but would vary relative to the benchmark initial margin by the extent of the price dynamics for the conditional distribution of futures.

This means that variation margins should be modelled conditionally (so that they can take account of current market dynamics) and at more conventional (i.e., non-extreme) confidence levels. The need to take account of current market conditions suggests that we should use some sort of GARCH process (e.g., as in Barone-Adesi *et al.* (1999), McNeil and Frey (2000), Giannopoulos and Tunaru (2005) and Cotter (2006)). We want conventional confidence levels because the clearinghouse is concerned about the prospect of possible default in the near future, but the confidence levels should not be too low because that would involve very frequent changes in variation margins, and this would be difficult to implement in practice.

In the literature margins have been typically modelled as a quantile or VaR.[4] The clearinghouse then selects a particular confidence level, and sets the margin as the VaR at this confidence level.[5] However, the VaR has been heavily criticised as a risk measure as it does not satisfy the properties of coherence and, most particularly, because the VaR is not subadditive (Artzner *et al.* (1999), Acerbi (2004)).[6]

This paper examines how one may obtain variation margins using a conditional modelling framework, but using risk measures that are superior to the VaR. The model used is a AR(1)-GARCH(1,1), and the other risk measures considered are the Expected Shortfall (ES), which is the average of the worst $p$ losses, where $p$ is the tail probability or 1 minus the confidence level, and Spectral Risk Measures (SRMs), which are risk measures that take account of the user's (in this case, the clearinghouse's) degrees of risk aversion. Both these types of risk measure are coherent and, from a theoretical point of view, demonstrably superior to the VaR.

---

[4] An exception is Bates and Craine (1999) who estimate a measure for the expected additional funds required assuming the futures price move has exhausted the initial margin. These additional funds fund however would not just include the variation margins but would come from a variety of sources including the remaining assets of losing futures traders, the remaining assets of the Clearinghouse and possibly fund arising from central bank intervention.

[5] The Clearinghouse has a difficult balancing act for the margin system introduced between minimising counterparty risk and remaining competitive for trades: too high a margin implies low counterparty risk but also an uncompetitive environment for futures traders and vice versa.

[6] The failure of VaR to be subadditive can then lead to strange and undesirable outcomes: in the present case, the use of the VaR to set margin requirements takes no account of the magnitude of possible losses exceeding VaR, and can therefore leave the clearinghouse heavily exposed to very high losses exceeding the VaR. For instance, because VaR is not-subadditve, using the VaR to set margin requirements might encourage investors to break up their accounts to reduce overall margin requirements, and in so doing leave the clearinghouse exposed to a hidden residual risk against which the clearinghouse has no effective collateral from its investors.

This paper provides estimates of conditional VaR, ES and spectral risk measures for long positions in 6 index futures contracts: S&P500, the FTSE100, the DAX, the Hang Seng, and the Nikkei225 indexes. The adequacy of the fits is confirmed by the results of a variety of backtests. The estimated risk measures illustrate the prevailing price dynamics of the conditional distributions of the futures. The paper also evaluates the precision of these estimates using a number of different ways of estimating their precision.

This paper is organised as follows. Section 2 reviews the risk measures to be examined and explores their properties. Section 3 looks at conditional risk modelling where returns follow an AR(1)-GARCH(1,1) process. Section 4 discusses data issues and model estimation, and section 5 evaluates the fitted model. Sections 6, 7 and 8 present results for the VaR, ES and SRM risk measures in turn. Section 9 discusses our findings and section 10 concludes. The paper is then followed by two appendices, one discussing the parametric bootstrap used in the simulations and the other discussing how numerical integration methods may be used to estimate Spectral Risk Measures.

## 2. MEASURES OF RISK

We are interested in risk measures that are weighted-averages of the quantiles of the return distribution. If $q_p$ is the 100$p$% quantile of the return distribution, then we can specify our risk measure $M_\phi$ as:

$$M_\phi = -\int_0^1 \phi(p) q_p dp \qquad (1)$$

for some arbitrary weighting function $\phi(p)$. The specification of $\phi(p)$ then determines the risk measure itself.

The best-known risk measure in this class is the Value-at-Risk (VaR). The VaR at the 100$\alpha$% confidence level is equal to $-q_{1-\alpha}$, so the risk measure (1) is equivalent to the VaR when $\phi(p)$ takes the form of a Dirac delta function that gives the $q_{1-\alpha}$ quantile an infinite weight (such that $\int_0^1 \phi(p)dp = 1$) and gives every other quantile a weight of zero. This means that that the VaR weighting function places no weight on tail quantiles, and implies that a VaR user is (in some sense) held to 'care more' about the prospect of a loss equal to VaR than about the prospect of a loss greater than the VaR. This is of course a rather strange property and one that leads the VaR to be non-subadditive[7] and therefore not coherent (see Acerbi (2004, p. 174); see also Artzner et alia (1999)).[8]

A second risk measure is the Expected Shortfall (ES), which is the average of the worst $(1-\alpha)100\%$ of losses, i.e.:

$$ES_\alpha = -\frac{1}{1-\alpha}\int_\alpha^1 q_p dp \qquad (2)$$

The ES is therefore based on a weighting function that gives each tail quantile the same weight, and gives every other quantile a weight of zero:

$$\phi(p) = \begin{cases} \frac{1}{1-\alpha} \\ 0 \end{cases} \text{ if } p = \begin{cases} p \geq \alpha \\ p < \alpha \end{cases} \qquad (3)$$

---

[7] Let $A$ and $B$ represent any two portfolios, and let $\rho(.)$ be a measure of risk over a given horizon. The risk measure $\rho(.)$ is subadditive if it satisfies $\rho(A+B) \leq \rho(A) + \rho(B)$. Subadditivity reflects risk diversification and is the most important a priori criterion we would expect a 'respectable' risk measure to satisfy. However, it can be demonstrated that VaR is not subadditive unless we impose the empirically implausible requirement that returns are elliptically distributed. Its non-subadditivity makes it very difficult to regard the VaR as a 'respectable' measure of risk.

[8] There are also other reasons to believe that VaR does not sit well with 'well-behaved' utility or risk aversion functions. For example, the VaR is not consistent with expected utility maximisation except in the very unusual case where risk preferences are lexicographic (Grootveld and Hallerbach, 2004, p. 33). Another example emerges from the downside risk literature (see, e.g., Bawa (1975) and Fishburn (1977)), which suggests that we can think of downside risk in terms of lower-partial moments (LPMs), where the LPM of order $k \geq 0$ around a below-target return $r*$ is equal to $E[\max(0, r*-r)^k]$. The parameter $k$ reflects the degree of risk aversion, and the user is risk-averse if $k > 1$, risk-neutral if $k = 1$, and risk-loving if $0 < k < 1$. From the LPM perspective, we would choose the VaR as our preferred risk measure only if $k = 0$ (Grootveld and Hallerbach, 2004, p. 35), and this suggests that the use of the VaR as a preferred risk measure indicates a strong degree of *negative* risk aversion.

Unlike the VaR, the ES has the attraction of being a coherent risk measure. However, since the ES weighting function gives the same weight to all tail quantiles, the choice of the ES as a risk measure would imply that the user is risk-neutral at the margin between better and worse tail outcomes, and this is inconsistent with risk-aversion.[9] Furthermore, like the VaR, the ES is conditioned on a parameter, the confidence level, whose value is typically difficult to establish and is often selected arbitrarily.

If we wish to have a risk measure that takes account of user risk-aversion, we can use a spectral risk measure in which $\phi(p)$ is obtained from the user's risk-aversion function. In practice, this requires that we specify the form this function takes, but a plausible choice is an exponential risk-aversion function which implies the following weighting function:

$$\phi(p) = \frac{ke^{-kp}}{1-e^{-k}} \qquad (4)$$

where $k \in (0,\infty)$ is the user's coefficient of absolute risk-aversion (see Acerbi (2004, p. 178)) or Cotter and Dowd (2006)). This function depends on a single conditioning parameter, the coefficient of absolute risk aversion, the value of which reflects the risk aversion of the user. A spectral risk-aversion function is illustrated in Figure 1. This shows how the weights rise as we encounter the prospect of higher losses and the rate of increase depends on $k$: the more risk-averse the user, the more rapidly the weights rise.

Insert Figure 1 here

Unlike the case with the VaR or ES confidence level, the value of the risk-aversion parameter is in principle known or at least ascertainable, and this means that spectral risk measures avoid the conditioning parameter arbitrariness implicit in using the other two risk measures. Given the risk-aversion function, setting the value of the user's risk-aversion parameter ensures that the spectral risk measure takes a unique value:

$$M_\phi = -\int_0^1 \phi(p) q_p dp = -\frac{k}{1-e^{-k}} \int_0^1 e^{-kp} q_p dp \tag{5}$$

which would typically be estimated using some form of numerical integration or quadrature method (e.g., a trapezoidal rule, Simpson's rule, pseudo or quasi Monte Carlo, etc.).[10] The fact that an SRM takes account of the user's attitude to risk also means that an SRM is a subjective risk measure in a way that the VaR or ES are not: so two users with the same portfolio but differing degrees of risk-aversion would face SRMs of different values, but still face the same VaR or ES.

---

[9] Again from the downside risk literature the ES is the ideal risk measure if $k=1$, implying that the user is risk-neutral (Grootveld and Hallerbach, 2004, p. 36).

As well as reflecting user risk-aversion, a spectral risk measure is also coherent provided $\phi(p)$ is nonnegative for all $p \in [0,1]$, provided $\int_0^1 \phi(p) dp = 1$ and provided $\phi(p)$ satisfies the weakly increasing property of giving higher losses weights that are no smaller than those of lower losses. In terms of Figure 1, this latter property means that $\phi(p)$ must never fall as we go from left to right along the *x*-axis (Acerbi (2004, proposition 3.4) This property indicates that the key to coherence is that a risk measure must give higher losses at least the same weight in (1) as lower losses. This also helps explain why the ES is coherent and why the VaR is not, and tells us that the VaR's most prominent inadequacies are closely related to its failure to satisfy the weakly increasing property (see also Acerbi (2004, p. 173)).

## 3. MODELLING CONDITIONAL RISK

Following McNeil and Frey (2000) and Cotter (2006), we model the daily return process by a normal AR(1)-GARCH(1,1) process. This process supposes that daily returns $r_t$ are conditionally normal, i.e.,

---

[10] More details on such methods can be found in standard references (e.g., Kreyszig (1999, pp. 869-878) or Miranda and Fackler (2002, chapter 5).

$$r_t \sim N(\mu_t, \sigma_t^2) \tag{6}$$

where the mean $\mu_t$ obeys an AR(1) process:

$$\mu_t = E[r_t] = \rho r_{t-1} + \varepsilon_t \quad ; \quad |\rho| < 1 \tag{7}$$

and $\varepsilon_t$ is iid zero-mean normal, and where the variance $\sigma_t^2$ obeys a GARCH(1,1) process:

$$\sigma_t^2 = \omega + \alpha r_{t-1}^2 + \beta \sigma_{t-1}^2 \quad ; \quad \omega \geq 0 \; , \; \alpha, \beta \geq 0 \; , \; \alpha + \beta < 1 \tag{8}$$

The AR(1)-GARCH(1,1) is a popular and parsimonious model which often provides a reasonable fit to daily return data. This model allows daily returns to have some degree of persistence, to have a volatility that exhibits persistence but also alternates between periods of low and high volatility, and to have moderate degrees of skewness and excess kurtosis.

Now let $M_{\phi,t}(\mu_t, \sigma_t)$ be our time-varying forecast of a risk measure for day $t$. Note that this is written as a function of the forecasts of the day-$t$ mean and volatility, $\mu_t$ and $\sigma_t$, and these two parameters are sufficient to calibrate the forecast of the risk measure forecast because we have assumed that returns are conditionally normal. For any of the three types of risk measure – VaR, ES or SRM – it is easy to show that the following relationship holds:

$$M_{\phi,t}(\mu_t, \sigma_t) = -\mu_t + \sigma_t M_{\phi,t}(0,1) \tag{9}$$

This relationship is handy computationally, because it reduces the task of risk forecasting to the two very simple smaller tasks – calculating the standard normal risk measure $M_{\phi,t}(0,1)$, and forecasting the parameters $\mu_t$ and $\sigma_t$ - and this can sometimes lead to major gains in computational efficiency.[11] Once we specify the relevant conditioning parameter, then the first task becomes straightforward and we can focus on the second. To illustrate, suppose that for the VaR and the ES we set the confidence level $\alpha$ to be 95%. And if the risk measure is the VaR, then the standard normal risk measure $M_{\phi,t}(0,1)$ becomes:

$$M_{\phi,t}(0,1) = z_\alpha = z_{0.95} = 1.6449 \qquad (10)$$
$$\Rightarrow M_{\phi,t}(\mu_t, \sigma_t) = -\mu_t + \sigma_t \times 1.6449$$

If the risk measure is the ES, then

$$M_{\phi,t}(0,1) = \frac{\phi(z_\alpha)}{1-\alpha} = \frac{\phi(z_{0.95})}{0.05} = 2.0627 \qquad (11)$$
$$\Rightarrow M_{\phi,t}(\mu_t, \sigma_t) = -\mu_t + \sigma_t \times 2.0627$$

where $\phi(.)$ is the value of the standard normal density. On the other hand, if the risk measure is an SRM and we take the ARA coefficient to be, say, 50, then $M_{\phi,t}(0,1)$ becomes:

$$M_{\phi,t}(0,1) = \frac{k}{1-e^{-k}} \int_0^1 e^{-kp} z_p \, dp \approx 2.2403 \qquad (12)$$
$$\Rightarrow M_{\phi,t}(\mu_t, \sigma_t) = -\mu_t + \sigma_t \times 2.2403$$

In each case risk forecasting now boils down to the task of forecasting $\mu_t$ and $\sigma_t$.

Table 1 gives a set of alternative values of $M_{\phi,t}(0,1)$ for different risk measures and conditioning parameters. So, for instance, if one was interested in the ES at the 99% confidence level, one would use (11) with $M_{\phi,t}(0,1) = 2.6652$ and the only information then needed to estimate the ES would be forecasts of the parameters $\mu_t$ and $\sigma_t$. Similarly, if one was interested in the SRM with ARA=100, one would use (12) with $M_{\phi,t}(0,1) = 2.4916$, and so forth.

---

[11] This issue is discussed further in Appendix 1.

Insert Table 1 here

## 4. DATA AND MODEL ESTIMATION

Our data set consists of daily geometric returns (taken as the difference between the natural logarithms of consecutive end-of-day prices) for the most heavily traded index futures – that is, the S&P500, FTSE100, DAX, Hang Seng and Nikkei 225 futures – between January 1 2000 and December 31 2002. Daily price data was obtained from Datastream representing the full calendar period excluding weekends giving 782 close-of-day returns. This sample period is then split into two subsamples: an initial estimation period (covering all of 2000 and 2001) to provide an initial GARCH fit; and a rolling estimation period (encompassing all of 2002) over which the model is updated on a daily basis.

As a preliminary, Table 2 provides some descriptive statistics for the full sample period are outlined in Table 2. The mean returns are near zero but negative, and the standard deviations of returns are in excess of 1% per day for all indexes. Most returns have a small negative skew, and all returns exhibit moderate degrees of excess kurtosis. The Ljung-Box statistics applied to the return series give a mixed picture – two of them appear to have a significant dependence structure, and the other three do not. These results suggest that we might wish to take account of possible dependence in the return process, and are the reason why we chose to model returns using an AR(1) process. For their part, the Ljung-Box statistics applied to squared returns indicate that these have very significant dependence structures, and this finding is reinforced by the significant ARCH effects indicated by the results of Engle's (1981) LM test. These latter test results support the existence of time-varying volatility dynamics and suggest that some form of GARCH-type process is called for.

Insert Table 2 here

The AR(1)-GARCH(1,1) was first applied to each futures index daily returns for 2000-2001. Estimation was by Maximum Likelihood. We then obtain forecasts for each week day in 2002 by updating parameters on a daily rolling window basis giving 259 sets of time-varying conditional parameters.

Table 3 reports the average values of the parameters and of their associated diagnostics. Our results are very much as expected: GARCH effects are significant and parameter values across the different indices are similar to each other and to those reported by many other studies.[12] The residual diagnostic results also suggest that the residuals are independent and that the model is well-specified.

Insert Table 3 here

Figure 3 shows plots of the estimated GARCH daily volatilities in each of our five futures indices over 2002.[13] The volatility of the S&P is a little more than 1% for most of the first half of 2002, but then it rises afterwards and twice peaks at around 2.5%. The FTSE volatility is similar, but peaks at greater values and is noticeably higher in the second half of the year. The DAX volatilities show a similar pattern but are a little higher than the FTSE ones. The Asian markets are much more tranquil: the Hang Seng volatility is generally a little under 1.5% in early 2002 and then rises a little quite different for most of the second half of the year, but is always well under 2%; the Nikkei is mostly between 1% and 2% and peaks a little beyond 2% in early March, but then falls back again is stable for the rest of the year. Thus, Figure 3 shows that conditions varied somewhat across different markets, and perhaps the most noticeable difference is that western markets exhibited much less tranquillity in later 2002 than did their east Asian counterparts.

Insert Figure 2 here

---

[12] We present average values only for brevity, but there was relatively little fluctuation in parameter estimates on a day-to-day basis. Further summary statistics details (deviations etc) are available on request.

[13] We only show results for long positions because there is little asymmetry in the index returns: explicitly addressing the VaRs of short positions would therefore provide little substantial additional information.

**5. MODEL EVALUATION**

In order to provide a more formal evaluation of the model, we must first set out its main predictions. One useful prediction relates to the model's probability integral transform (PIT) values. The PITs are the values of the realised returns after they have been put through the following transformation:

$$p_t = F_t(r_t) \qquad (13)$$

where $F_t(.)$ is the forecasted cumulative density function made the previous day. A sample of PIT values is predicted to be distributed as standard uniform under the null hypothesis of model adequacy.[14] Accordingly, Diebold *et alia* (1998, p. 869) suggest that a useful diagnostic of model adequacy is to plot the PITs and check visually if they are 'close' to the predicted uniform distribution. The PIT values from our model applied to the futures returns are presented in Figure 3. The fitted lines are very close to the 45° line predicted under the null, and this strongly suggests that the fits are good ones.

Insert Figure 3 here

Another prediction of the model is that the frequency of exceedances – that is, the frequency of observed losses exceeding VaR – should be compatible with the predictions of the model. This lead to the Kupiec test: for the 95% VaR, the model predicts that 5% of observations should be exceedances. With 259 observations, a predicted 5% exceedance probability means that there are $259 \times 0.05 \approx 13$ predicted exceedances, so we would test the hypothesis that the number of exceedances is acceptably close to 13. We can test this prediction using a binomial test.

---

[14] A formal proof of this prediction is provided by Diebold *et alia* (1998, pp. 865, 867-869).

A third prediction relates to the standardized residuals. Combining and rearranging (6) and (7) leads to the prediction that the standardized residuals $\varepsilon_t/\sigma_t$ should be iid N(0,1), viz.:

$$\varepsilon_t/\sigma_t \sim \text{iid N}(0,1) \qquad (14)$$

Thus, $\varepsilon_t/\sigma_t$ are predicted to be both iid and standard normal. However, the results of Table 3 have already established that $\varepsilon_t$ are independent, so we can work on the basis of that the iid prediction is satisfied. We therefore focus on textbook tests of standard normality taking iid for granted. These include:

- a *z*-test of the prediction that $\varepsilon_t/\sigma_t$ has a mean of zero;
- a *t*-test of the prediction that $\varepsilon_t/\sigma_t$ has a mean of zero;
- a variance-ratio test of the prediction that $\varepsilon_t/\sigma_t$ has a unit variance; and
- a Jarque-Bera test of the prediction that $\varepsilon_t/\sigma_t$ has zero skewness and a kurtosis of 3.

The results of these tests are presented in Table 4. For each futures index, this Table presents the number of exceedances, the mean, standard deviation, skewness and kurtosis of $\varepsilon_t/\sigma_t$, and the prob-values of each of the above tests. Going through these results one index at a time:

- The S&P does well by all tests except perhaps the Kupiec one, which yield a prob-value of 3.7%. This latter result suggests that the number of exceedances (19) might be rather high, which might indicate that the model under-estimates the 95% VaR. The FTSE results are similar, except that the prob-value of the Kupiec test is now 0.003. This presents strong evidence that the number of exceedances (23) is too high, and that the model under-estimates the 95% VaR.
- The DAX does well with the possible except of the *z*-test and *t*-test, which are both significant at under the 5% level. This might indicate a problem with the prediction that standardised residuals should have a zero mean.

- The Hang Seng does well for all tests except the Jarque-Bera. The Jarque-Bera is significant at under the 1% level, and this suggests that the standardized residuals are not normal.
- The Nikkei passes all tests easily.

In sum, of 25 test results, 20 are not significant, 3 are significant at the 5% level and 2 are significant at the 1% level. Overall, we would suggest that this is a reasonably good performance which suggests that the model's forecasts are fairly accurate.

Insert Table 4 here

**6. RESULTS FOR VALUE-AT-RISK**

Figure 4 shows plots of the estimated 95% VaRs and the estimated bounds of their 90% confidence intervals for long positions in each of our five futures indices over 2002. Broadly speaking, the estimated VaRs show much the same patterns as the GARCH volatilities and paint a similar picture about market conditions. For their part, the confidence bounds in Figure 4 suggest that the uncertainty in VaR forecasts had a tendency to move with the VaR forecasts themselves. This tendency is particularly pronounced with the FSTE, where the confidence bounds are initially quite narrow but become much wider when the forecasted VaR peaks in August and then again in October. We see comparable increases in the widths of the S&P and DAX confidence bounds when they also peak at much the same times.

Insert Figure 4 here

Table 5 gives VaR results computed for the 2002 average daily values of the input parameters. The Table reports estimated VaRs and a variety of precision metrics for our VaR estimates: these are the VaR standard error (SE); the standardized VaR SE, which is the SE divided by the estimated VaR; the 90% confidence interval for the VaR; and the standardized 90% confidence interval for the VaR, which is same confidence interval divided by the estimated VaR. The first and the third of these give estimates of precision in absolute terms, whereas the second and fourth give estimates of precision relative to the size of the estimated VaR.

These precision metrics are based on a fully parametric bootstrap applied to the AR(1)-GARCH(1,1) process. This bootstrap gives estimates of precision metrics based on simulated VaRs (or other risk measures) based on information available the previous day. Since the volatility follows a GARCH process, this means that the current-day volatility is already known (from (8)). Hence it is important to appreciate that the only noise in the bootstrap process is the sampling variation of the daily mean, and this means that the noisiness of the simulated VaRs is driven entirely off the noise in the simulated means. Further details of the bootstrap are given in Appendix 1.

Before examining the results, it is also important to appreciate that if $\sigma_t$ is given in any bootstrap simulations (as is the case here), then (9) tells us that the standard error of any of the risk measures considered here must be equal to the standard deviation of the $\mu_t$. This also means that, for any given sets of parameters, the VaR, ES and SRM must all have the same SE. We would emphasise that these predictions are not a product of any 'strangeness' in our algorithm, because the algorithm fully reflects the structure of the AR(1)-GARC(1,1) process and our assumptions about the information available at any point in time. Instead, these predictions are generated by the underlying structure of the model itself.

The Table shows that the VaRs are generally quite close. The S&P, FTSE, Hang Seng and Nikkei are quite close in the region 2.3% to 2.6%. The Hang Seng lowest at 2.338%, and the DAX is an outlier at 3.557%. The other three are close to the Hang Seng, so the DAX is an outlier. The Table reports that the VaR SEs are proportional to the VaRs, and the standardized SEs are all the same at 0.303%. The Table also shows that the bounds of the confidence interval reflect the sizes of the estimated VaRs, but the standardized confidence intervals – the confidence intervals divided by the estimated VaRs – are the same: the standardized 90% confidence interval is always [0.501,1.499], i.e., plus or minus 50% of the estimated VaR. The 'precision story' is therefore that in absolute terms, the level of precision varies inversely with the size of the VaR, but in relative or standardized terms, the level of precision is always the same.

Insert Table 5 here

## 7. RESULTS FOR EXPECTED SHORTFALL

The corresponding ES results are presented in Table 6. Perhaps the most striking feature in this Table is that the estimated ES is always approximately 25.5% greater than the comparable earlier VaR. This suggests that whatever the values of $\mu_t$ and $\sigma_t$ might be, the normal ES estimated with these parameter values will be close to 1.255 times the normal VaR estimated with these same parameter values.[15] It is therefore not surprising to discover that the ES plots in Figure 5 have the same shape as the early VaR ones, and therefore have the same interpretations. However, we would emphasise that this relationship is approximate rather than exact: if we plot the ratio of the ES to the VaR over time, we do not get a straight line, but noisy process that oscillates around a straight line and is close to it.

Insert Table 6 here
Insert Figure 5 here

The precision metrics in Table 6 are also as expected: the ES has the same SE as the VaR, and therefore has a small standardized SE. The bounds of the ES confidence level are then pushed out by the extent of the difference between the ES and the VaR, but the ES confidence interval has the same width as the VaR one. And because the ES exceeds the ES, the ES must have a narrower standardised SE than the VaR: the standardized ES confidence interval is now plus or minus 40%. Thus, in absolute terms, the ES is estimated with the same precision as the VaR, but in relative terms, it is estimated more precisely than the VaR. And, if we compare the ES results to each other, results show that in absolute terms precision varies inversely with the size of the estimated ES, but in relative terms, the level of precision is always the same.

---

[15] For any given confidence level and given empirically plausible parameter values, the normal ES is a slightly varying greater-than-one multiple of the normal VaR. The multiple itself can be seen in the ratio of the standard normal ES to the standard normal VaR. So, for example, with a confidence level of 95% the ratio of the standard normal ES to the standard normal VaR can be seen from Table 1 to be 2.0627/1.6449=1.2540≈1.2555 as in Table 6.

## 8. RESULTS FOR SPECTRAL RISK MEASURES

The corresponding SRM results are presented in Table 7.[16] These show that the SRM calibrated on an ARA coefficient of 50 is now in the region of 1.361 times the value of its 95% VaR counterpart,[17] and this implies (and Figure 6 confirms) that the SRM plots have much the same shapes as the VaR ones in Figure 3. Of course, we should keep in mind that whereas the normal ES always exceeds the normal VaR predicated on a given confidence level, the normal SRM predicated on a chosen ARA does not always exceed the normal VaR predicated on a given confidence level: if the ARA is relatively low, and the confidence level high, then the SRM can be lower than the VaR (as is evident from Table 1).

Insert Table 7 here
Insert Figure 6 here

The precision metrics in Table 7 are also as expected: the SRM has the same SE as the VaR, and therefore has a smaller standardized SE. The bounds of the SRM confidence level are pushed out by the difference between the SRM and the VaR, and the SRM confidence interval has the same width as the VaR one; and the standardized SRM confidence interval is now plus or minus 36%. In absolute terms, the SRM is estimated with the same precision as the VaR, but in relative terms, it is estimated more precisely. (However, for reasons that will be apparent from the previous paragraph, we would expect the SRM to be relatively less precisely estimated than the VaR in cases where the ES is smaller rather than larger than the VaR.) Comparing the SRM results to each other, we find (as with the VAR and ES) that absolute precision varies inversely with the size of the estimated risk measure, but in relative terms, the level of precision is always the same.

---

[16] More details on the estimation of the SRMs are provided in Appendix 2.

[17] Reminiscent of the last note but one, the SRM predicated on a particular ARA coefficient is a slightly varying but not-necessarily-greater-than-one multiple of the α VaR, where the multiple can again be inferred from the results of Table 1. In this case, Table 1 tells us that the standard normal SRM predicated on an ARA of 50 is equal to 2.2376, and the 95% standard normal VaR is 1.6449. The ratio 2.2376/1.6449=1.360≈1.361, which is the value reported in Table 7.

## 9. DISCUSSION

For all three risk measures, we therefore get the same 'precision story': in absolute terms precision varies inversely with the size of the estimated risk measure, but in relative terms, the level of precision is always the same.

Second, all risk measures indicate the relative riskiness of different contracts and how the risk changes over time. For instance, each of the conditional risk measures show the time varying nature of volatility for the respective indices during 2002. All indices exhibit specific dynamics reflecting high conditional volatility at a certain time and thereby having large associated risk estimates followed by decreasing conditional volatility towards the end of the sample as indicated by the decreasing risk estimates. However as all the risk estimates are driven by the same conditional process the same pattern emerges for each risk measure. Clearly they all suggest that the Hang Seng contract is the least risky index and that the DAX is the most risky. Thus, the use of any of these measures for setting variation margins would therefore lead to the former ones having the lowest margins and the DAX the highest.

Third, both estimates of ES and VaR are directly comparable to each other as their key parameter is the confidence level, unlike the SRM, whose key parameter is the degree of risk aversion.

Moreover there are also distinctions in terms of the interpretation and usefulness of the risk measures. First, the use of VaR to estimate variation margins only allows the Clearinghouse to estimate a quantile and the associated default probability. Thus, the conditional VaR is limited as it gives the Clearinghouse no idea of the size of their exposure beyond the probability level chosen. However, estimating a conditional VaR (and the other measures) allows the Clearinghouse to estimate their variation margins based on the time-varying dynamics prevalent at any time.

Second, in principle the ES is more useful to the clearinghouse than the VaR because it takes account of the sizes of losses higher than the VaR itself. It also has the helpful interpretation that it tells the clearinghouse the loss an investor can expect to make on it experiencing a loss that exceeds a chosen VaR threshold. So if the clearinghouse sets a VaR-based variation margin, then the ES tells the clearinghouse the expected default loss for the investor experiencing a loss that exceeds its margin. Also, by estimating a conditional ES the clearinghouse has an indication of the expected losses assuming a quantile has been breached based on the prevailing price dynamics.

Third, in setting variation margins, the spectral risk measures are in principle the most useful, because they alone take account of the user's degree of risk aversion.

## 10. SUMMARY AND CONCLUSIONS

Clearinghouses manage counterparty risk through a margin account. In addition to the initial margin, a trader must add a variation margin if a margin call takes place. The modelling of variation margins encompasses the price dynamics of the conditional distribution during the lifetime of the futures contract. This paper estimates variation margins using an AR(1)-GARCH (1, 1) process that models both the conditional mean and volatility using three different risk measures: the VaR, the Expected Shortfall (ES), and spectral-coherent risk measures. Although the time-varying approach allows for a daily update of variation margins based on each risk measure, the Clearinghouse may have variation margins that reflect the prevailing dynamics but updated less frequently, thereby exploiting the volatility persistence in futures.

Previous studies have estimated margins for futures returns using a VaR measure where margins are associated with quantiles measured at confidence levels. In contrast, the use of ES allows the Clearinghouse to get an estimate of the expected losses assuming the margin level is breached. Moreover, the use of SRMs allows the Clearinghouse to incorporate risk aversion into the margin estimates. This paper illustrates the properties and the associated relative merits of the measures underpinned by the conditional distribution of futures.

**REFERENCES**

Acerbi, C., 2002. Spectral measures of risk: a coherent representation of subjective risk aversion. Journal of Banking and Finance 26: 1505-1518.

Acerbi, C., 2004. Coherent representations of subjective risk-aversion, in G. Szego (Ed), Risk Measures for the 21$^{st}$ Century, Wiley, New York, pp. 147-207.

Artzner, P., F. Delbaen, J.-M. Eber, D. Heath, 1999. Coherent measures of risk. Mathematical Finance 9, 203-228.

Barone-Adesi, G., K. Giannopoulos, and L. Vosper, 1999. VaR without correlations for portfolios of derivatives securities, Journal of Futures Markets, 19, 583-602.

Bates, D., and R. Craine, 1999. Valuing the futures market clearinghouse's default exposure during the 1987 crash. Journal of Money, Credit, and Banking, 31, 248-272.

Bawa, V. S., 1975. Optimal rules for ordering uncertain prospects. Journal of Financial Economics 2: 95-121.

Bollerslev, T., Chou, R., and K. Kroner, 1992. ARCH modelling in finance: a review of the theory and empirical evidence, Journal of Econometrics, 52, 5-59.

Booth, G. G., Brousssard, J.P., Martikainen, T., and Puttonen, V., 1997. Prudent margin levels in the Finnish stock index futures market. Management Science 43, 1177-1188.

Brennan, M.J., 1986. A theory of price limits in futures markets, Journal of Financial Economics, 16, 213-233.

Cotter, J. 2001. Margin exceedences for European stock index futures using extreme value theory. Journal of Banking and Finance 25, 1475-1502.

Cotter, J., 2006, Varying the VaR for unconditional and conditional environments, Journal of International Money and Finance, Forthcoming.

Cotter, J., Dowd, K., 2006, Extreme spectral risk measures: an application to futures, Electronically available on Science Direct: http://www.sciencedirect.com/science?_ob=ArticleListURL&_method=list&_ArticleListID=477617306&_sort=d&view=c&_acct=C000050221&_version=1&_urlVersion=0&_userid=10&md5=e64db1df9db6e25e4c1394873d9aeb0d
20

## APPENDIX 1: AN EFFICIENT PARAMETRIC BOOTSTRAP

The precision metrics (i.e., standard errors and confidence intervals) reported in the paper were estimated using a parametric bootstrap: this bootstrap is motivated by the idea that our precision metrics we should make full use of the structure of the model. Now imagine the following problem. We would like to estimate precision metrics for a risk measure $M_{\phi,t}(\mu_t, \sigma_t)$ for day *t*, using information available for day *t*-1. To do so, we need a bootstrap procedure that gives us a set of, say, *m* randomly chosen values of $M_{\phi,t}(\mu_t, \sigma_t)$, where these simulated values make use of the information we have about day *t*-1 and are generated in a way that reflects the AR(1)-GARCH(1,1) model, i.e., we want a fully parametric bootstrap.

The logic behind this bootstrap is as follows: we first note from (9) that we can construct a value of $M_{\phi,t}(\mu_t, \sigma_t)$ if we have values of the parameters $\mu_t$ and $\sigma_t$. We also know from the GARCH(1,1) process (8) that $\sigma_t$ is determined by our information from the previous day, i.e., $\sigma_t$ is already given. Any randomly simulated value of $M_{\phi,t}(\mu_t, \sigma_t)$ must then be driven by a randomly simulated value of $\mu_t$. This means that we need to simulate values of $\mu_t$ and (7) tells us that $\mu_t$ is a zero-mean normal with an unknown variance. Let us suppose for the moment that we know this variance. We now simulate a value of $\mu_t$ from this normal distribution and input this value and our given value of $\sigma_t$ into (9) to simulate a random value of the risk measure $M_{\phi,t}(\mu_t, \sigma_t)$. We then do this a large number of times *m* to give us a large sample of simulated $M_{\phi,t}(\mu_t, \sigma_t)$ values.

Our precision metrics then follow naturally: if we want the standard error of $M_{\phi,t}(\mu_t, \sigma_t)$, we estimate this as the standard deviation of $M_{\phi,t}(\mu_t, \sigma_t)$; and if we want a confidence interval, we can use a conventional textbook formula.

It remains to show how we obtain the variance of $\mu_t$. The easiest way to obtain this numerically is to first assume an arbitrary variance for $\mu_t$ and use this in (6) to simulate a set of returns. (We can actually use the same set of simulated random numbers as before, but this is just an issue of computational efficiency.) We then estimate the variance of these returns and find this to be, say, $\sigma_t^*$. However, the returns should have a variance of $\sigma_t$. To get the correct variance for both returns and $\mu_t$, we therefore have to multiply the original assumed variance by $\sigma_t/\sigma_t^*$. The 'true' value of the variance of $\mu_t$ is then equal to $\sigma_t/\sigma_t^*$ times the original assumed variance.

We should note two other points about this bootstrap. First, the random noise in our simulated $M_{\phi,t}(\mu_t,\sigma_t)$ values is driven off the random noise in the $\mu_t$ process. Since (9) also tells us that $M_{\phi,t}(\mu_t,\sigma_t)$ moves pari passu with $\mu_t$, this implies that the standard error of $M_{\phi,t}(\mu_t,\sigma_t)$ must be equal to the volatility of $\mu_t$. This latter prediction also means that for any given set of parameters, the VaR, ES and SRM all have the same standard error, and the results in Tables 5 to 7 reflect these implications.

The final point to note is that we use (9) in our algorithm in order to ensure that we only ever 'directly' estimate the standard normal risk measure $M_{\phi,t}(0,1_t)$, and that we do this only once in each bootstrap exercise; all estimates of $M_{\phi,t}(\mu_t,\sigma_t)$ are then obtained 'indrectly' by inserting the standard normal estimate $M_{\phi,t}(0,1_t)$ and the parameters $\mu_t$ and $\sigma_t$ into (9). This is much more efficient computationally than estimating $M_{\phi,t}(\mu_t,\sigma_t)$ directly (i.e., by inputting parameter values into the relevant version of (1)). So, for example, if we are dealing with an SRM, the former approach requires us to invoke a numerical integration routine only once, but the latter requires us to invoke it a large number of times. The former approach is much faster computationally. Thus, our bootstrap algorithm is computationally efficient as well as fully parametric.

# APPENDIX 2 : ESTIMATING SPECTRAL RISK MEASURES USING NUMERICAL INTEGRATION

Unlike the case with the estimation of VaR or ES, the estimation of spectral risk measures typically requires us to compute the value of an integral. Where returns are normal with mean $\mu_t$ and standard deviation $\sigma_t$, then equations (9) and (12) tell us that the SRM predicated on a coefficient of absolute risk aversion equal to *k* is given by:

$$M_{\phi,t}(\mu_t,\sigma_t) = -\mu_t + \sigma_t M_{\phi,t}(0,1) \tag{A2.1}$$

where

$$M_{\phi,t}(0,1) = -\frac{k}{1-e^{-k}} \int_0^1 e^{-kp} z_p \, dp \tag{A2.2}$$

where $z_p$ is the 100*p*% quantile of the standard normal distribution. We therefore have to calculate the integral in (A2.2). We can do so using a numerical integration or quadrature method, in which we approximate the continuous integral by a discrete equivalent: we discretise the continuous variable *p* into a number *N* of discrete slices, where the approximation gets better as *N* gets larger. To apply numerical intergration, we have to select a value of *N* and choose a suitable numerical integration method, and our choices include trapezoidal and Simpson's rules, quasi-Monte Carlo methods (e.g., such as those using Niederreiter or Weyl algorithms) and pseudo-Monte Carlo methods.

To evaluate the accuracy of these methods, Figure A2.1 shows plots of estimated standard normal SRMs against *N*, where the SRM is predicated on an ARA coefficient of 50, using the trapezoidal rule, Simpson's rule, Niederreiter and Weyl quasi-Monte Carlo, and pseudo-Monte Carlo numerical integration methods. We consider *N* values from 100 to 50000. The first four converge upwards towards the 'true' value of 2.2403 and are all more or less accurate when *N* reaches 50000. By contrast, the pseudo-Monte Carlo gives very volatile estimates, and converges much more slowly. We can therefore eliminate the pseudo-Monte Carlo method as very unreliable compared to the others. Of the remaining four, the trapezoidal and Simpson's rule estimates are smoother and converge somewhat faster than the quasi-Monte Carlo methods. There is very little difference between the trapezoidal and Simpson's rule estimates, and both are very accurate for *N* values of 20000 or more. However, the trapezoidal rule is easier computationally. Taking on board issues of both accuracy and computational efficiency, we then selected the trapezoidal rule with *N*=30000 as the combination of *N* value and integration method to be used to produce the SRM estimates in the paper: this combination produces highly accurate results fairly quickly.

Insert Figure A2.1 here

**TABLES**

| Table 1: Values of Standard Normal Risk Measures ||||| 
|---|---|---|---|---|
| $\alpha$ | VaR | ES | ARA | SRM |
| 0.75 | 0.6745 | 1.2711 | 1 | 0.2779 |
| 0.8 | 0.8416 | 1.3998 | 5 | 1.0809 |
| 0.85 | 1.0364 | 1.5544 | 10 | 1.5031 |
| 0.9 | 1.2816 | 1.7550 | 15 | 1.7139 |
| 0.925 | 1.4395 | 1.8874 | 20 | 1.8509 |
| 0.95 | 1.6449 | 2.0627 | 25 | 1.9514 |
| 0.975 | 1.9600 | 2.3378 | 50 | 2.2376 |
| 0.99 | 2.3263 | 2.6652 | 100 | 2.4916 |
| 0.995 | 2.5758 | 2.8919 | 500 | 2.9671 |

Notes: 'VaR', 'ES' and 'SRM' are Value-at-Risk, Expected Shortfall and Spectral Risk Measures for standard normally distributed returns, '$\alpha$' is the confidence level and 'ARA' is the coefficient of absolute risk aversion. The SRM measures are estimated using the trapezoidal numerical integration method with $N$=30000.

| Table 2: Summary Statistics for Daily Return Data ||||||
|---|---|---|---|---|---|
| | S&P | FTSE | DAX | HANG SENG | NIKKEI |
| Mean | -0.067 | -0.074 | -0.113 | -0.077 | -0.101 |
| Std Dev | 1.474 | 1.440 | 1.898 | 1.790 | 1.616 |
| Skewness | 0.148 | -0.121 | -0.004 | -0.126 | 0.143 |
| Kurtosis | 4.296 | 4.590 | 4.495 | 4.998 | 4.593 |
| Minimum | -6.271 | -6.062 | -9.243 | -8.712 | -7.599 |
| Maximum | 5.755 | 5.350 | 7.289 | 6.431 | 8.004 |
| Q(12) | 33.706 | 62.075** | 49.811* | 31.279 | 32.897 |
| Q(12)$^2$ | 168.122** | 712.201** | 561.477** | 95.414** | 74.666** |
| ARCH(12) | 92.935** | 190.816** | 162.925** | 60.997** | 60.726** |

Notes: Based on the 782 close-of-day % returns for each of the stated indexes over the period January 1 2000 to December 31 2002. Mean, standard deviation, minimum and maximum are in percentage form. Q(12) is the 12-lag Ljung-Box test statistic applied to the returns series, $Q^2(12)$ is the same test statistic applied to the squared returns series, and ARCH(12) is the Engle (1981) LM test for up to twelve-order ARCH effects. * represent significant at 5% level and ** represents significance at 1% level.

## Table 3: Mean GARCH Parameter Estimates and Diagnostics for Futures Indexes

|  | S&P500 | FTSE100 | DAX | Hang Seng | Nikkei 225 |
|---|---|---|---|---|---|
| $\rho$ | -0.058 | -0.046 | -0.057 | -0.066 | -0.101 |
|  | (0.592) | (0.362) | (0.411) | (0.383) | (0.179) |
| $\bar{\omega}_t$ | 0.089 | 0.067 | 0.098 | 0.105 | 0.161 |
|  | (0.097) | (0.025) | (0.098) | (0.105) | (0.066) |
| $\bar{\alpha}_t$ | 0.083 | 0.128 | 0.108 | 0.059 | 0.059 |
|  | (0.001) | (0.000) | (0.000) | (0.010) | (0.011) |
| $\bar{\beta}_t$ | 0.872 | 0.829 | 0.859 | 0.905 | 0.883 |
|  | (0.000) | (0.000) | (0.000) | (0.000) | (0.000) |
| Q(12) – R | 29.461 | 61.032 | 42.412 | 36.229 | 27.893 |
|  | (0.339) | (0.004) | (0.061) | (0.134) | (0.452) |
| $Q(12)^2$ – R | 68.020 | 339.280 | 230.680 | 74.101 | 79.639 |
|  | (0.000) | (0.000) | (0.000) | (0.003) | (0.001) |
| ARCH(12) - R | 45.039 | 110.406 | 107.643 | 54.560 | 61.669 |
|  | (0.016) | (0.000) | (0.000) | (0.018) | (0.006) |
| AIC | 1770.164 | 1651.647 | 1976.771 | 2034.182 | 2001.613 |
| BIC | 1787.187 | 1668.676 | 1993.800 | 2051.210 | 2018.643 |
| JB-Z | (45.039) | 13.829 | 5.635 | 43.234 | 39.573 |
|  | (0.016) | (0.006) | (0.130) | (0.001) | (0.004) |
| Q(12) – Z | 13.330 | 13.617 | 11.749 | 15.473 | 6.071 |
|  | (0.345) | (0.349) | (0.471) | (0.266) | (0.883) |
| $Q(12)^2$ – Z | 14.020 | 5.611 | 10.634 | 4.297 | 20.213 |
|  | (0.299) | (0.927) | (0.562) | (0.970) | (0.190) |
| ARCH(12) - Z | 13.270 | 6.057 | 10.442 | 4.428 | 20.952 |
|  | (0.350) | (0.901) | (0.579) | (0.968) | (0.173) |

Notes: The Table presents average GARCH model coefficients and average diagnostics. Marginal significance levels using Bollerslev-Wooldridge standard errors are displayed in parentheses. Pre-model diagnostics are applied to the returns series R and post model diagnostics are applied to the standardised filtered return series Z. Q(12) is the Ljung-Box test applied to the indicated series, and $Q^2(12)$ is the Ljung-Box test applied to the squared indicated series. ARCH(12) is the Engle (1981) LM test for up to twelfth order ARCH effects. Marginal significance levels for the model diagnostics are given in parentheses.

### Table 4: Model Evaluation Results

|  | S&P | FTSE | DAX | Hang Seng | Nikkei |
|---|---|---|---|---|---|
|  | Numbers of exceedances | | | | |
|  | 19 | 23 | 18 | 11 | 9 |
|  | Summary statistics for standardized residuals | | | | |
| Mean | -0.0981 | -0.1036 | -0.128 | -0.0632 | -0.0546 |
| Std | 1.0368 | 1.0174 | 1.0312 | 0.9437 | 0.9639 |
| Skewness | 0.1017 | -0.1199 | 0.0797 | 0.1849 | 0.2848 |
| Kurtosis | 3.2819 | 3.0305 | 2.7114 | 3.9219 | 3.1070 |
| Test | Prob-values of test statistics | | | | |
| Kupiec | 0.0373* | 0.0030** | 0.0629 | 0.3531 | 0.1624 |
| $z$-test | 0.1144 | 0.0954 | 0.0394* | 0.3088 | 0.3800 |
| $t$-test | 0.1291 | 0.1024 | 0.0468* | 0.2818 | 0.3632 |
| Variance ratio | 0.3896 | 0.6723 | 0.4630 | 0.2073 | 0.4264 |
| Jarque-Bera | 0.5209 | 0.7297 | 0.5562 | 0.0049** | 0.1633 |

*Notes*: The results apply to 259 daily observations over 2002. The number of exceedances refers to the number of loss returns exceeding forecasted VaR. The standardized residuals are the residuals of (6) divided by forecasted $\sigma_t$. The tests are two-sided version of: the Kupiec binomial test that the frequency of exceedances equals 1 minus the confidence level of 95%, $z$- and $t$-tests of the prediction that standardized residuals have a mean equal to 0, a variance-ratio test of the prediction that standardized residuals have a variance equal to 1, and a Jarque-Bera test that the standardized residuals are normal. * indicates significance at 5% level and ** indicates significance at 1% level.

### Table 5: Value-at-Risk Results

|  | VaR | SE | St. SE | 90% CI | | St. 90% CI | |
|---|---|---|---|---|---|---|---|
|  |  |  |  | LB | UB | LB | UB |
| S&P | 2.491 | 0.759 | 0.303 | 1.254 | 3.752 | 0.501 | 1.499 |
| FTSE | 2.512 | 0.765 | 0.303 | 1.265 | 3.782 | 0.501 | 1.499 |
| DAX | 3.557 | 1.085 | 0.304 | 1.791 | 5.361 | 0.501 | 1.499 |
| HS | 2.338 | 0.712 | 0.303 | 1.177 | 3.519 | 0.501 | 1.499 |
| Nikkei | 2.605 | 0.794 | 0.303 | 1.312 | 3.924 | 0.501 | 1.499 |

*Notes*: Estimates of VaR are in daily % return terms and VaR is predicated on a 95% confidence level. 'SE' is the VaR standard error, 'St. SE' is the standard error of the standardized VaR (i.e., VaR SE divided by the mean VaR), '90% CI' is the 90% confidence interval for the VaR, 'St. 90% CI' is the 90% confidence interval for the standardized VaR, and 'LB' and 'UB' refer to the lower and upper bounds of confidence intervals. Standard errors and confidence intervals are estimated using a parametric bootstrap applied to the 2002-average values of the 259 sets of daily parameters of the AR(1)-GARCH(1,1) model. To facilitate comparability of results across Tables 5 to 7, simulations are carried out using the same sets of seed numbers for the pseudo-random number generator.

## Table 6: Expected Shortfall Results

### (a) Results in comparable form to those of Table 5

|         | ES    | SE    | St. SE | 90% CI LB | 90% CI UB | St. 90% CI LB | St. 90% CI UB |
|---------|-------|-------|--------|-----------|-----------|---------------|---------------|
| S&P     | 3.126 | 0.759 | 0.242  | 1.889     | 4.387     | 0.602         | 1.398         |
| FTSE    | 3.151 | 0.765 | 0.242  | 1.904     | 4.421     | 0.602         | 1.398         |
| DAX     | 4.464 | 1.085 | 0.242  | 2.698     | 6.268     | 0.602         | 1.398         |
| HS      | 2.933 | 0.712 | 0.242  | 1.772     | 4.115     | 0.602         | 1.398         |
| Nikkei  | 3.269 | 0.794 | 0.242  | 1.975     | 4.587     | 0.602         | 1.398         |

### (b) Above results divided by those of Table 5

|         | ES    | SE    | St. SE | 90% CI LB | 90% CI UB | St. 90% CI LB | St. 90% CI UB |
|---------|-------|-------|--------|-----------|-----------|---------------|---------------|
| S&P     | 1.255 | 1.000 | 0.798  | 1.506     | 1.169     | 1.201         | 0.933         |
| FTSE    | 1.255 | 1.000 | 0.798  | 1.506     | 1.169     | 1.201         | 0.933         |
| DAX     | 1.255 | 1.000 | 0.798  | 1.506     | 1.169     | 1.202         | 0.933         |
| HS      | 1.255 | 1.000 | 0.798  | 1.506     | 1.169     | 1.201         | 0.933         |
| Nikkei  | 1.255 | 1.000 | 0.798  | 1.506     | 1.169     | 1.201         | 0.933         |

*Notes*: Estimates of ES are in daily % return terms and ES is predicated on a 95% confidence level. 'SE' is the ES standard error, 'St. SE' is the standard error of the standardized ES (i.e., ES SE divided by the mean ES), '90% CI' is the 90% confidence interval for the ES, 'St. 90% CI' is the 90% confidence interval for the standardized ES, and 'LB' and 'UB' refer to the lower and upper bounds of confidence intervals. Standard errors and confidence intervals are estimated using a parametric bootstrap applied to the 2002-average values of the 259 sets of daily parameters of the AR(1)-GARCH(1,1) model. To facilitate comparability of results across Tables 5 to 7, simulations are carried out using the same sets of seed numbers for the pseudo-random number generator.

## Table 7: Spectral Risk Measure Shortfall Results

### (a) Results in comparable form to those of Table 5

|         | SRM   | SE    | St. SE | 90% CI LB | 90% CI UB | St. 90% CI LB | St. 90% CI UB |
|---------|-------|-------|--------|-----------|-----------|---------------|---------------|
| S&P     | 3.391 | 0.759 | 0.223  | 2.154     | 4.652     | 0.633         | 1.367         |
| FTSE    | 3.419 | 0.765 | 0.223  | 2.172     | 4.689     | 0.633         | 1.367         |
| DAX     | 4.843 | 1.085 | 0.223  | 3.077     | 6.647     | 0.633         | 1.367         |
| HS      | 3.182 | 0.712 | 0.223  | 2.021     | 4.364     | 0.633         | 1.367         |
| Nikkei  | 3.546 | 0.794 | 0.223  | 2.253     | 4.865     | 0.633         | 1.367         |

### (b) Above results divided by those of Table 5

|         | SRM   | SE    | St. SE | 90% CI LB | 90% CI UB | St. 90% CI LB | St. 90% CI UB |
|---------|-------|-------|--------|-----------|-----------|---------------|---------------|
| S&P     | 1.361 | 1.000 | 0.736  | 1.718     | 1.240     | 1.263         | 0.912         |
| FTSE    | 1.361 | 1.000 | 0.736  | 1.717     | 1.240     | 1.263         | 0.912         |
| DAX     | 1.362 | 1.000 | 0.735  | 1.718     | 1.240     | 1.264         | 0.912         |
| HS      | 1.361 | 1.000 | 0.736  | 1.717     | 1.240     | 1.263         | 0.912         |
| Nikkei  | 1.361 | 1.000 | 0.736  | 1.718     | 1.240     | 1.263         | 0.912         |

*Notes*: Estimates of the spectral risk measure (SRM) are in daily % return terms and the SRM is predicated on a coefficient of absolute risk aversion equal to 50. SRMs are estimated using the Miranda-Fackler 'CompEcon' trapezoidal numerical integration MATLAB program with the number of slices equal to 30000. 'SE' is the SRM standard error, 'St. SE' is the standard error of the standardized SRM (i.e., SRM SE divided by the mean SRM), '90% CI' is the 90% confidence interval for the SRM, 'St. 90% CI' is the 90% confidence interval for the standardized SRM, and 'LB' and 'UB' refer to the lower and upper bounds of confidence intervals. Standard errors and confidence intervals are estimated using a parametric bootstrap applied to the 2002-average values of the 259 sets of daily parameters of the AR(1)-GARCH(1,1) model. To facilitate comparability of results across Tables 5 to 7, simulations are carried out using the same sets of seed numbers for the pseudo-random number generator.

**FIGURES**

**Figure 1: Spectral Loss Weighting Functions**

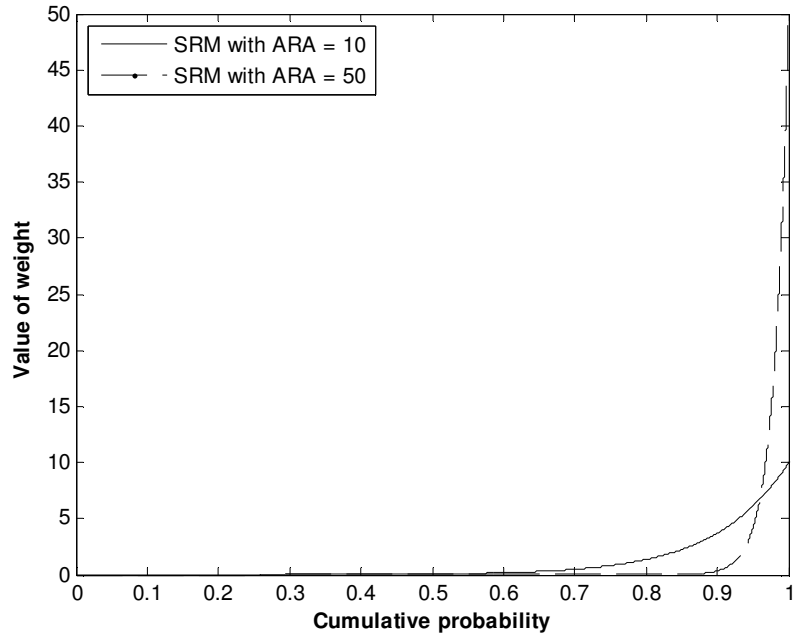

Notes: 'SRM' is the spectral risk measure based on equation (4) in the text, predicated on coefficients of absolute risk aversion (ARA) equal to 10 and 50.

**Figure 2: Plots of Daily GARCH (1,1) Volatilities**

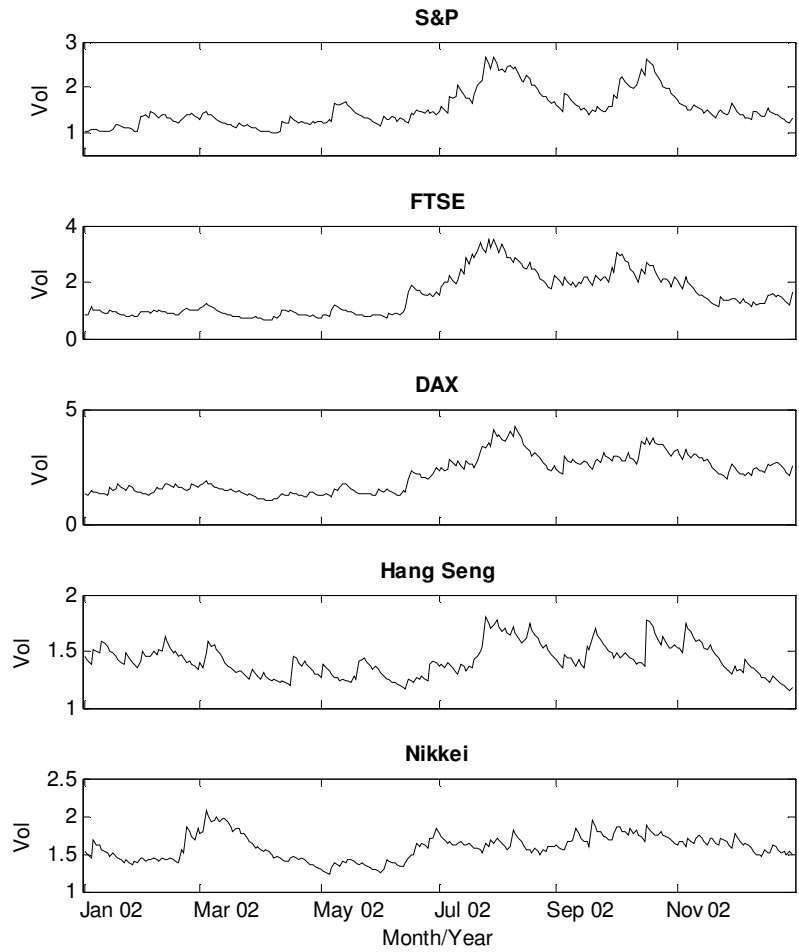

Notes: Volatilities are based of GARCH(1,1) process in daily % return terms. Results based on 259 daily observations over 2002.

**Figure 3 Probability Integral Transform Plots of Fitted AR(1)-GARCH(1,1) Process**

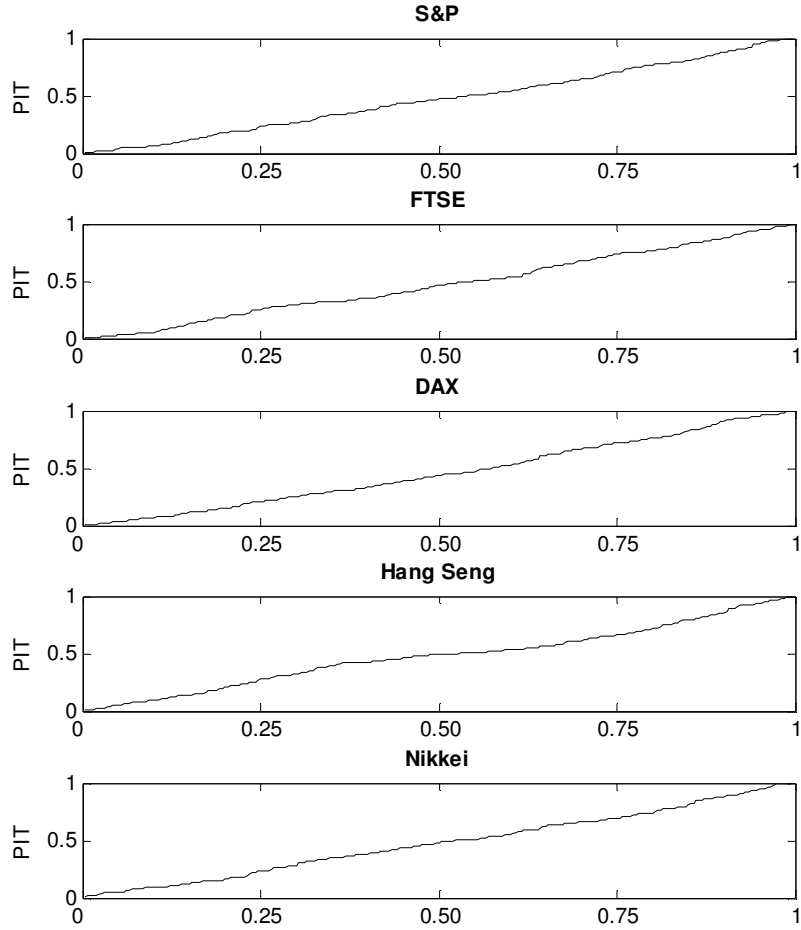

Notes: Plots show Probability Integral Transform (PIT) values for forecasts over each of 259 trading days in 2002 plotted against the unit interval [0,1].

**Figure 4: Plots of Daily AR(1)-GARCH (1,1) 95% VaR**

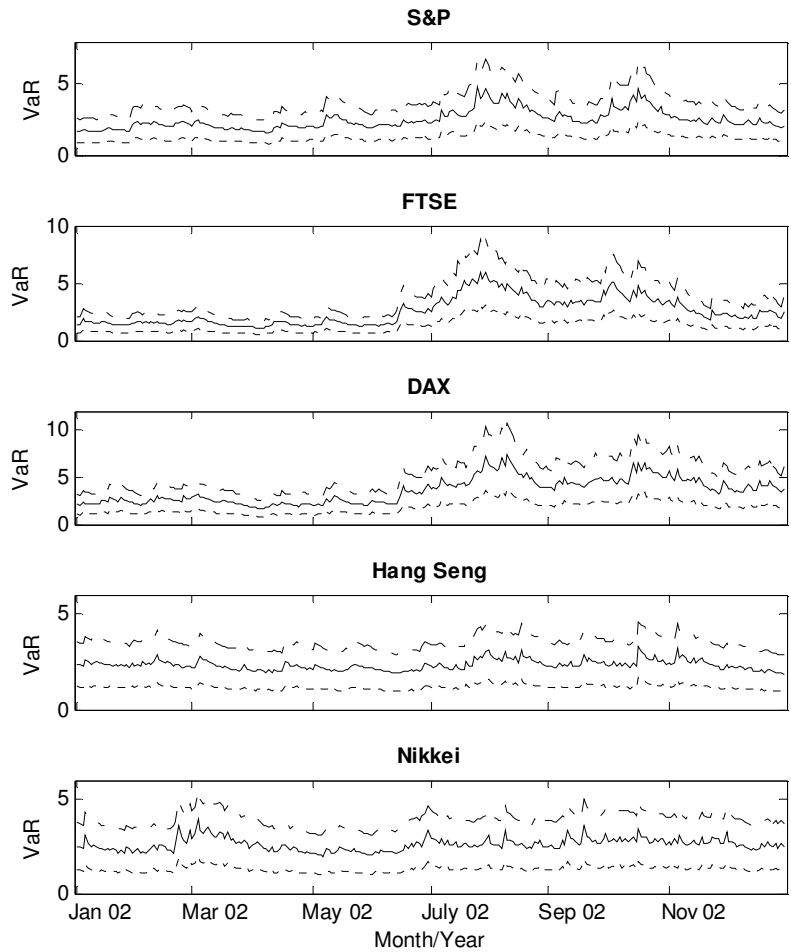

Notes: The VaR is that of a long position in the relevant index. Estimates of VaR are in daily % return terms and VaR is predicated on a 95% confidence level. Results based on 259 daily observations over 2002. Each plot shows estimated VaR plus estimated 90% confidence bounds for VaR.

**Figure 5: Plots of Daily AR(1)-GARCH (1,1) 95% Expected Shortfall**

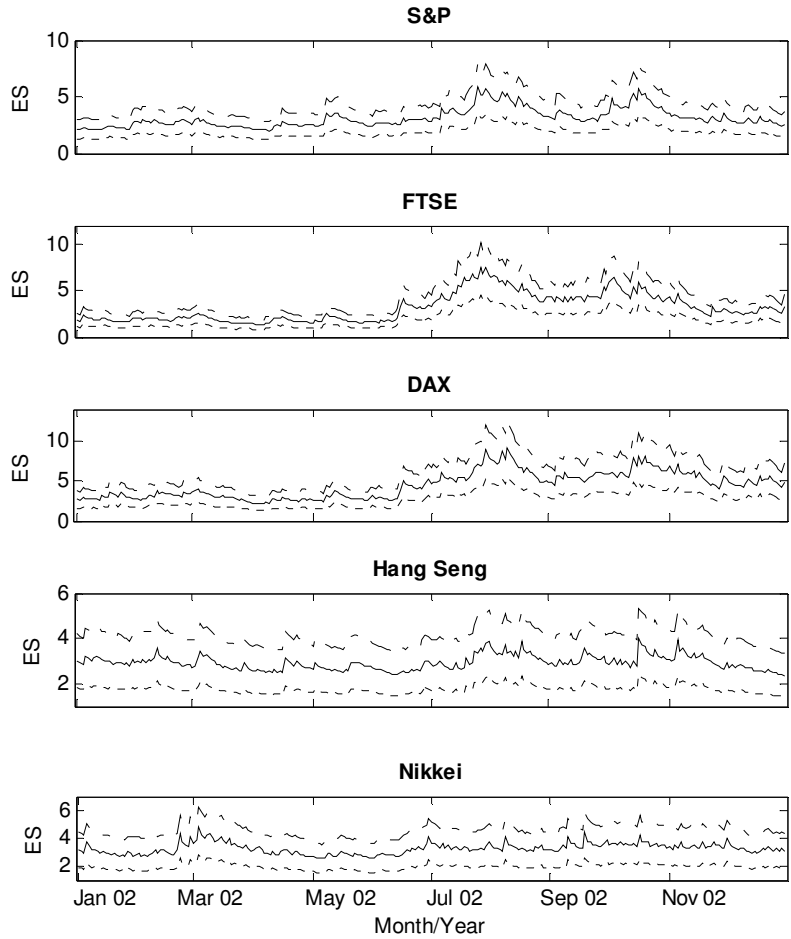

Notes: The ES is that of a long position in the relevant index. Estimates of ES are in daily % return terms and ES is predicated on a 95% confidence level. Results based on 259 daily observations over 2002. Each plot shows estimated ES plus estimated 90% confidence bounds for ES.

**Figure 6: Plots of Daily AR(1)-GARCH (1,1) Spectral Risk Measure**

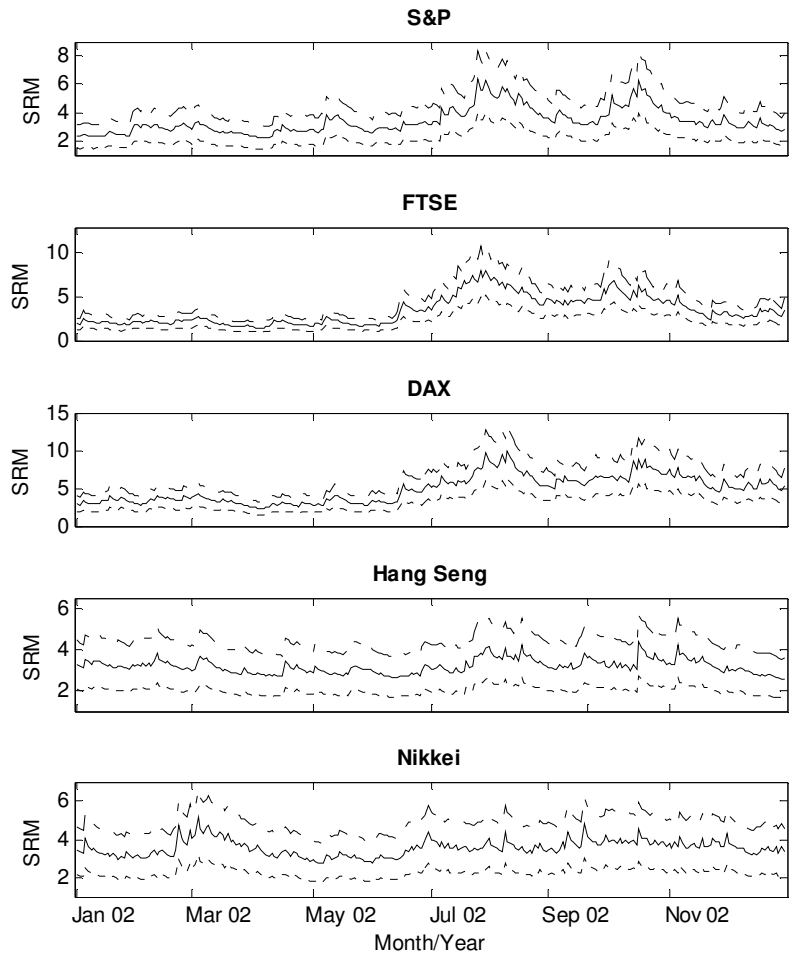

Notes: The SRM is that of a long position in the relevant index. Estimates of SRM are in daily % return terms and the SRM is predicated on a coefficient of absolute risk aversion equal to 50. SRMs are estimated using a trapezoidal numerical integration method with the number of slices equal to 30000. Results based on 259 daily observations over 2002. Each plot shows estimated SRM plus estimated 90% confidence bounds of SRM.

**Figure A2.1: Plots of Estimated Standard Normal Risk Measures Against the Number of Slices, *N***

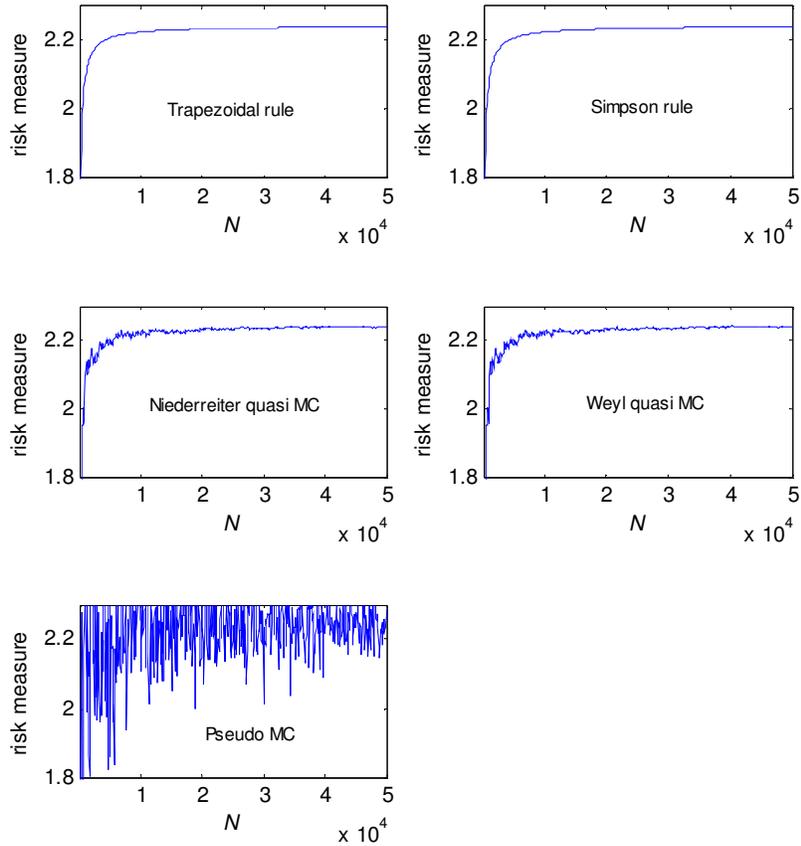

Notes: Each plot shows the estimated spectral-exponential risk measure against *N*, where *N* covers the range 100 to 50000 in steps of 100, obtained using the numerical integration routines shown on each plot. The spectral risk measure is predicated on a coefficient of absolute risk aversion equal to 50, and returns are assumed to be distributed as standard normal. Calcuations were carried out using the Miranda-Fackler 'CompEcon' numerical integration programs written in MATLAB.